# Nanoscale visualization of the thermally-driven evolution of antiferromagnetic domains in FeTe thin films

Shrinkhala Sharma[1], Hong Li[1], Zheng Ren[1*], Wilber Alfaro Castro[1] and Ilija Zeljkovic[1†]

[1] Department of Physics, Boston College, 140 Commonwealth Ave, Chestnut Hill, MA 02467

*zr10@rice.edu

†ilija.zeljkovic@bc.edu

## Abstract

Antiferromagnetic order, being a ground state of a number of exotic quantum materials, is of immense interest both from the fundamental physics perspective and for driving potential technological applications. For a complete understanding of antiferromagnetism in materials, nanoscale visualization of antiferromagnetic domains, domain walls and their robustness to external perturbations is highly desirable. Here, we synthesize antiferromagnetic FeTe thin films using molecular beam epitaxy. We visualize local antiferromagnetic ordering and domain formation using spin-polarized scanning tunneling microscopy. From the atomically-resolved scanning tunneling microscopy topographs, we calculate local structural distortions to find a high correlation with the distribution of the antiferromagnetic order. This is consistent with the monoclinic structure in the antiferromagnetic state. Interestingly, we observe a substantial domain wall change by small temperature variations, unexpected for the low temperature changes used compared to the much higher antiferromagnetic ordering temperature of FeTe. This is in contrast to electronic nematic domains in the cousin FeSe multilayer films, where we find no electronic or structural change within the same temperature range. Our experiments provide the first atomic-scale imaging of perturbation-driven magnetic domain evolution simultaneous with the ensuing structural response of the system. The results reveal surprising thermally-driven modulations of antiferromagnetic domains in FeTe thin films well below the Neel temperature.

## Introduction

Antiferromagnetic (AF) ordering is an important presence in the phase diagrams of various correlated electron systems. For example, in cuprate high-temperature superconductors, the ground state of the undoped parent system is an AF Mott insulator [1,2], which becomes superconducting as the AF Mott insulating state is weakened by chemical doping. In Fe-based high-temperature superconductors, AF order has been widely observed in the parent systems [3,4], which may have a close relationship with the subsequent emergence of electronic nematicity and superconductivity [3,5]. More recently, it has been discovered that many kagome metals also host AF ordering, which coexists with flat bands, Dirac fermions and charge density waves [6–9].

While magnetic domains are nearly ubiquitous in ferromagnets due to uncompensated stray magnetic fields and the resulting interactions, domains in antiferromagnets are generally less frequent. Nevertheless, antiferromagnets still exhibit a tendency to form domains [10–13]. A domain state denotes an ordered state associated with the possible order parameter orientation or phase; a domain is then defined as a translationally-invariant region in a solid that can take one of the possible domain states. Probing the AF domain structure and domain wall formation is of immense importance for the understanding of fundamental phenomena and for driving potential technological applications. AF spintronics, for example, is proposed to utilize AF moments as carriers for information, and strongly relies on the robustness of the AF structure to external perturbations [14–16]. AF domain walls may also have different properties compared to the interior of the domains [17,18].

Resolving magnetic domain structures at the nanoscale has been challenging but it can bring exciting new insights [12,19,20]. Probes that are crucial in unveiling magnetic ordering in solids, such as neutron scattering, nuclear magnetic resonance, muon spin rotation and resonant inelastic x-ray scattering, average the signal from the entire beam-spot and therefore lack the capability to visualize domains at the atomic scale. Complementary to these, scanning probe microscopy can be a very useful tool to visualize domains down to the nanoscale. Here we focus on using spin-polarized scanning tunneling microscopy (SP-STM), which measures the local spin-polarized density of states, to image magnetic ordering at the nanoscale [21]. In particular, we investigate the AF domain formation in FeTe thin films synthesized by molecular beam epitaxy (MBE). We identify the double stripe antiferromagnetic order in FeTe, as well as the domain structures consisting of nanometer-scale domains. By extracting local structural distortion from the atomically-resolved STM topographs, we find a high correlation between the antisymmetric strain and the AF domains, which is consistent with the structural distortion accompanying the AF phase in bulk FeTe. Surprisingly, we discover substantial fluctuations of magnetic domains after thermal cycling up to relatively low temperatures compared to the Neel temperature of FeTe.

**Results**

While the superconducting FeSe does not exhibit any magnetic ordering, double-stripe antiferromagnetic order emerges when Se is completely substituted with Te, accompanied by stronger electron-electron correlation [3–5]. As such, FeTe is often viewed as a parent compound of the superconducting chalcogenide Fe(Se,Te). FeTe has the PbO-type crystal structure (Fig. 1a), and it goes through a bulk antiferromagnetic phase transition at $T_N \approx 70$ K, accompanied by a tetragonal-monoclinic structural transition [22]. In contrast to the (π, π) in-plane stripe antiferromagnetic order in Fe pnictides, the in-plane antiferromagnetic order in FeTe appears to be a (π, 0) order characterized by the double stripe structure [3–5,22,23].

We synthesize FeTe thin films on Nb-doped SrTiO$_3$(001) substrates by using MBE (Methods). The quality of our films is confirmed by the sharp streaks in the reflection high energy electron diffraction (RHEED) pattern of FeTe (Fig. 1b). X-ray diffraction measurement exhibits only the

(00$l$) ($l$=1,2,3,4) reflections, which further demonstrates the layered structure of our films that grow along the crystalline *c*-axis (Fig. 1d). Based on the zero-field cooled resistance measurements as a function of temperature, we determine the antiferromagnetic ordering temperature to be $T_N \approx 62$ K (Fig. 1c). This is comparable but slightly lower than that of bulk single crystals of FeTe [22]. The surface morphology of the FeTe films is displayed in the STM topograph acquired using a conventional (spin-averaged) tungsten tip (Fig. 1e,f). As expected from the RHEED measurement, we observe an atomically flat surface with terraces. A step height of about 0.7 nm is measured, which is consistent with the expected unit cell height of bulk single crystals (inset of Fig. 1e). Top-most Te atoms are clearly visible in the atomically resolved topograph (Fig. 1f).

To gain insight into the magnetic structure of the FeTe films, we perform SP-STM measurements using spin-polarized tips (Methods). The same technique has been applied to a variety of different antiferromagnets, including bulk single crystals of FeTe [23–25], iridates [10,11], and chiral [20] and kagome magnets [6,9]. We focus on the area as shown in Fig. 2a,b. The SP-STM topograph $T(\mathbf{r}, B = 4$ T), acquired when a 4 T magnetic field is applied parallel to the *c*-axis, resolves the top-most Te atoms arranged on a square lattice (Fig. 2a). In addition to the Te atoms, a stripe-like superstructure is discernable propagating along the Te-Te *x*-axis, which has a wavelength of $2a_{Te}$ (Fig. 2a,d). These stripes are not seen in STM topographs obtained by a spin-averaged STM tip (Fig. 1f) and are consistent with the orientation and periodicity of the double stripe antiferromagnetic ordering observed in FeTe bulk crystals and in previous SP-STM work [23–25]. To further confirm the AF origin of these features, we reverse the direction of the magnetic field. This flips the polarization direction of the STM tip, without significantly affecting the magnetic ordering in the FeTe film, since the Zeeman energy induced by the 4 T field is much smaller than the energy scale of the exchange interaction. We image the identical region of the sample to find similar stripe features in the SP-STM topograph $T(\mathbf{r}, B = -4$ T), but with one noticeable difference – the bright stripes now shift by $a_{Te}$ along the *x*-axis (Fig. 2b,e). We obtain a spin-resolved magnetic contrast $M(\mathbf{r})$ map by subtracting the STM topographs in Fig. 2a and Fig. 2b, which emphasizes the AF contrast and now more clearly shows the $2a_{Te}$ stripe modulation related to the AF order (Fig. 2c,f). The topographs in Fig. 2a,b were drift-corrected and aligned with atomic-registry using Lawler-Fujita drift-correction algorithm [26] prior to subtraction. This conclusively demonstrates that the stripe-like features in SP-STM data are intimately related to the same double stripe antiferromagnetic ordering observed in FeTe bulk crystal.

We turn to a larger field of view to investigate the existence of AF domains (Fig. 3). We again acquire the SP-STM topographs over an identical field-of-view at different magnetic fields, $T(\mathbf{r}, B=4$ T) and $T(\mathbf{r}, B=-4$ T), and subtract the two to extract the AF signal (Fig. 3a). The $M(\mathbf{r})$ map shows the AF ordering stripes oriented in either near-horizontal, or near-vertical direction. The two types of orientational AF domains are enclosed by solid white lines, across which the orientation of the AF ordering stripes rotates by 90° (Fig. 3a). Within each orientational domain, we also find several smaller sub-domains with the same wave vector, but offset by π phase with respect to one another (denoted by dashed white lines in Fig. 3a, Supplementary Figure 7).

In FeTe bulk single crystals, it has been established that the antiferromagnetic phase transition is accompanied by a tetragonal-to-monoclinic structural transition. The in-plane lattice constants along the two lattice directions are different by about 1% in the AF phase [27,28]. To measure if the AF order in our FeTe film is also accompanied by a similar structural change, we use the Lawler-Fujita drift-correction algorithm to calculate the antisymmetric strain map $U(\mathbf{r})=u_{xx}(\mathbf{r})-u_{yy}(\mathbf{r})$ ($u_{ii}(\mathbf{r})= du_i(\mathbf{r})/dr_i$, $i=x,y$, where $u_i(\mathbf{r})$ is the displacement field), which quantifies local structural anisotropy between different crystalline directions [26,29–33] (Fig. 3b, Supplementary Note 1). By visually comparing the spin-resolved magnetic contrast $M(\mathbf{r})$ map and the $U(\mathbf{r})$ map (Fig. 3a,b), it is evident that the shape of the antiferromagnetic domains is correlated with the distribution of the structural anisotropy. To quantitatively establish the cross-correlation between the strength of AF ordering and the local structural anisotropy, we Fourier-filter the $M(\mathbf{r})$ map by choosing the $\mathbf{q}_{AF}$ peak along the x-axis (Fig. 3c) or along the y-axis (Fig. 3d) in the FT. We then calculate the radially-averaged cross-correlation coefficient of the Fourier-filtered $M(\mathbf{r})$ map and the $U(\mathbf{r})$ map, and find it to be 0.5 and -0.37 (Fig. 3c,d insets). Consistent with the expectation from the bulk FeTe, the lattice constant along which $\mathbf{q}_{AF}$ develops is larger by about 1 % [22].

Magnetic domains we visualized in Figure 3 are formed spontaneously when the sample is cooled across the ordering temperature. Thermal cycling through the ordering temperature can reform the domain structure [34,35]. However, it is not expected that the antiferromagnetic domains would significantly change at a temperature much lower than the Neel temperature. To investigate the robustness of the AF domain structure, we warm up the sample to a higher temperature, cool it back down to the base temperature and compare the AF orders before and after the thermal cycle. In Figure 4, we show one such process, where we warmed up the sample to ~10 K and cooled it back down to ~4.5 K (additional data is shown in Supplementary Figure 4). Focusing on the same field-of-view, we repeat the SP-STM experiment sequence to generate spin-resolved magnetic contrast maps before ($M(\mathbf{r})$) and after ($M'(\mathbf{r})$) thermal cycling (Fig. 4c,f). Surprisingly, by comparing $M(\mathbf{r})$ and $M'(\mathbf{r})$ maps, we observe a notable change of the orientational AF domain structure. In particular, a substantial part of the AF domain where the stripes were initially aligned along the x-axis in the $M(\mathbf{r})$ map rotated by ~90° to merge with a part of the AF domain along the other stripe orientation (along the y-axis) in the $M'(\mathbf{r})$ map. We also observe a change in the anti-phase sub-domains within each orientational domain (Figure 4c,f, Supplementary Figure 7). The considerable change in the AF domain structure with thermal cycling in evident in subsequent thermal cycles over the same area of the sample (Supplementary Figure 4). In particular, we cycled the system to 21 K and to 52 K (two times in a row) to find a different resulting domain structure at low temperature each time (Supplementary Figure 4).

**Discussion**

We can rule out the possibility that the AF domain modulation is driven by structural degrees of freedom because 10 K is extremely low compared to the synthesis temperature of FeTe films (about 570 K). Furthermore, in a similar heterostructure FeSe/SrTiO$_3$(001), we do not observe any change in the structure of electronic nematic domains, which also locally carry a 1% lattice anisotropy, by warming up to 30 K (Supplementary Figure 6). Future experiments could explore if monolayer FeTe grown on SrTiO$_3$ also shows the surprising $T_N$ enhancement similar to that reported in ultrathin FeTe films grown on Bi$_2$Te$_3$ [36]. It is interesting to note that the irregularly-shaped AF domain walls observed in thin films are markedly different than the larger scale FeTe domains with straight-line domain walls in FeTe bulk single crystals [23–25,37,38]. This is similar to the electronic nematic domain morphology in non-magnetic FeSe, where the domains in thin films are similarly irregularly shaped [31,33,39,40] and smaller compared to bulk FeSe [41,42]. The susceptibility of AF domain structure to change with small temperature variations may suggest the domain walls are not strongly pinned to defects or impurities in this system. As our FeTe films also exhibit strain at the order of a few percent due to the lattice mismatch with the substrate (Methods), this points to an important role of the substrate in partitioning the domains to smaller length scales compared to bulk single crystals.

**Methods**

**MBE growth.** SrTiO$_3$ (100) substrates (5 mm x 5 mm x 0.5 mm) were cleaned in acetone and 2-propanol in an ultrasonic bath and then introduced into our MBE system (Fermion Instruments) with a base pressure of ~ 5 x 10$^{-10}$ Torr. Nb-doped (0.05 wt%) SrTiO$_3$ was used as the substrate for the growth of FeTe films subsequently studied by STM, and undoped SrTiO$_3$ was used as the substrate for FeTe films characterized by transport measurements in Fig. 1c. The substrate was first slowly heated to the growth temperature at ~300°C, which was continuously monitored by a pyrometer (emissivity=0.7). Thereafter, Fe (99%) and Te (99.9999%) were co-evaporated from individual Knudsen cells after the flux rates were calibrated using a quartz crystal microbalance (QCM). A flux ratio of Fe:Te=1:15 was roughly achieved as the temperatures of Fe and Te were set at 1100°C and 240°C, respectively. Films studied in this work were 24 nm thick, determined from Fe flux calibration by QCM. XRD measurements of our films in Fig. 1d reveal the average *c*-axis lattice constant reduction by 3.8%, and in turn suggest, on average, in-plane tensile strain. For STM measurements, thin films were quickly transferred using a vacuum suitcase chamber held at 5 x 10$^{-11}$ Torr, and were never exposed to air.

**STM measurements.** STM data was acquired using a customized Unisoku USM1300 STM at the base temperature of about 4.5 K. Spectroscopic measurements were acquired using a standard lock-in technique at 915 Hz and bias excitations as detailed in the figure captions. STM tips were home-made chemically etched tungsten tips, annealed in UHV to bright orange color before STM imaging. Spin polarization of STM tips was obtained in-situ on FeTe films, by fast scanning and bias pulsing, which likely leads to the tip picking up a few Fe atoms at its apex. To obtain spin-resolved magnetic contrast $M(\mathbf{r})$ maps, the two topographs obtained at different magnetic fields were drift-corrected and aligned using Lawler-Fujita drift-correction algorithm [26] prior to subtraction.

**Acknowledgements**

I.Z. acknowledges the support from U.S. Department of Energy (DOE) Early Career Award DE-SC0020130.

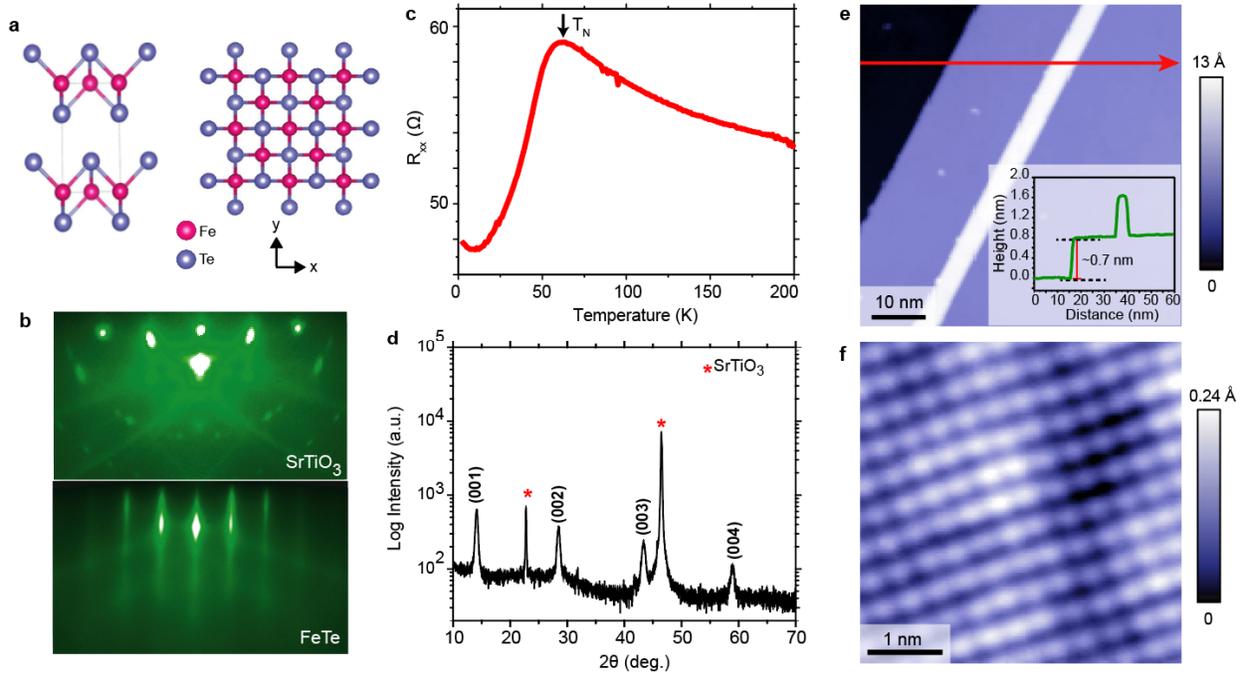

**Figure 1. Structural characterization of FeTe thin films grown on SrTiO$_3$.** (a) Crystal structure of bulk FeTe (left) and top view of the Te terminated surface and the underlying Fe lattice (right). Pink (blue) spheres denote the Fe (Te) atoms. (b) Reflection high energy electron diffraction (RHEED) pattern recorded before (top half) and after (both half) growth of 24 nm thick FeTe film on SrTiO$_3$. (c) Zero-field cooled resistance measurements as a function temperature showing a kink at the onset of the AF order. (d) X-ray diffraction data of the same thin film in (b) showing the high quality epitaxial growth. (e) STM topograph over a large region of the film. Inset shows a topographic line profile along the red line in (e). The step height is about 0.7 nm, consistent with the unit cell of FeTe. (f) Atomically resolved STM topograph showing the square lattice of Te atoms. STM setup condition: (e) $V_{sample}$ = 1V, $I_{set}$ = 10pA, 4.5 K; (f) $V_{sample}$ = 200 mV, $I_{set}$ = 800 pA, 4.5 K.

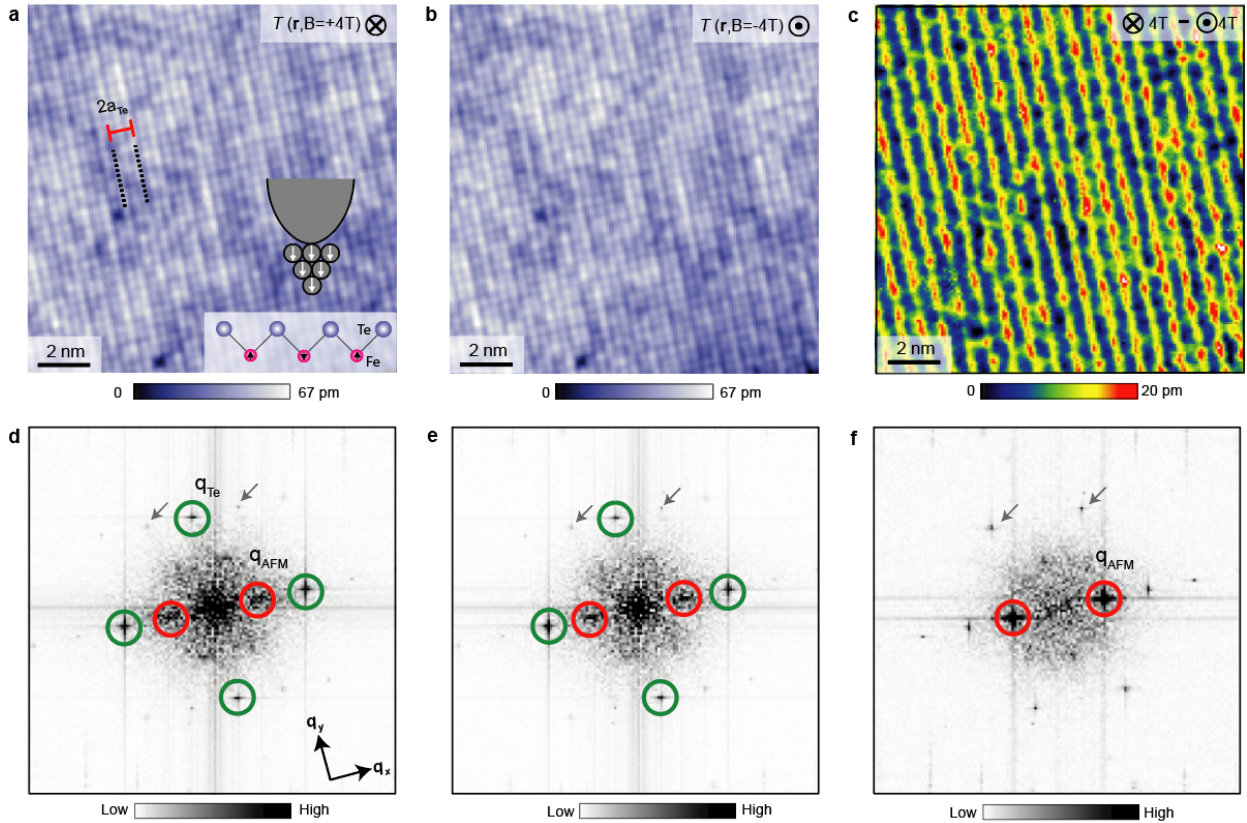

Figure 2. Spin-polarized scanning tunneling microscopy imaging of the double stripe antiferromagnetic (AF) order in 24 nm thick FeTe thin films grown on SrTiO$_3$. (a,b) STM topographs obtained using a spin-polarized tip at (a) +4 T and (b) -4 T magnetic field (the minus sign denotes reversal of the magnetic field applied parallel or antiparallel to the sample c-axis). Reversal of the magnetic field serves to flip the spin polarization of the tip. STM topographs show the surface Te atoms with additional $2a_{Te}$ periodic superstructure. The inset in (a) is a schematic illustration of the spin polarized STM measurement. (c) Spin-resolved magnetic contrast M(r) map obtained by subtracting imaged in (a) and (b). The map more clearly shows the bicollinear antiferromagnetic (AF) ordering. (d-f) Fourier Transforms of images in (a-c). Atomic Bragg peaks $q_{Te}$ are circled in green and the additional AF peaks $q_{AFM}$ = ½ $q_{Te}$ are circled in red. Small gray arrows in (d-f) denote the reflections of the AF peaks from the atomic Bragg peaks. The FeTe film is about 24 nm thick. STM setup condition: (a,b) $V_{sample}$ = 30mV, $I_{set}$ = 200pA, 4.5 K.

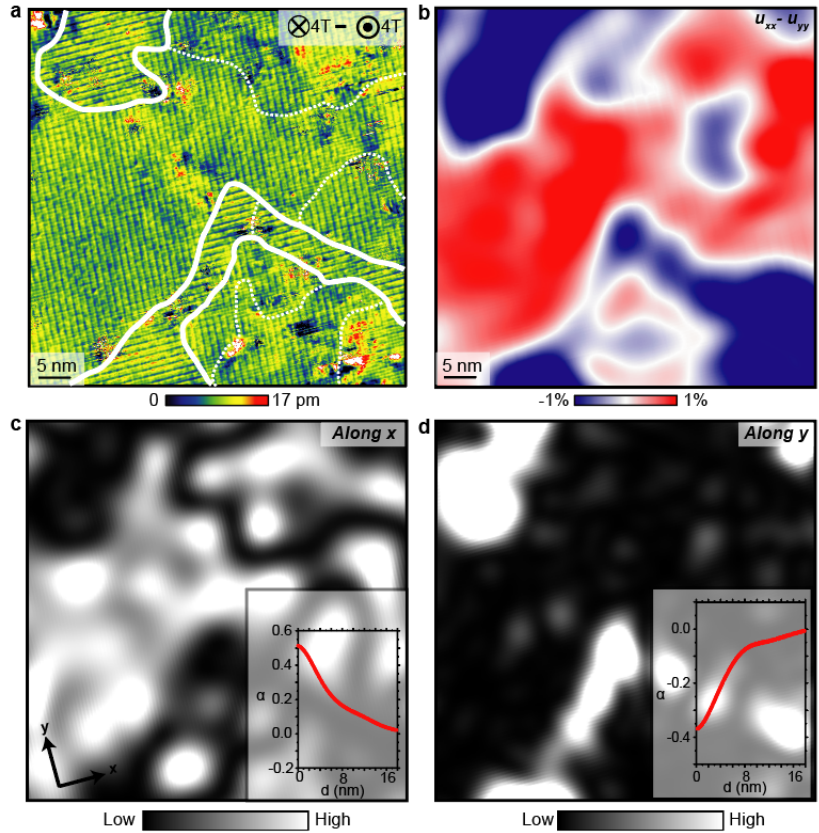

**Figure 3. Correlation of strain with antiferromagnetic order.** (a) Spin-resolved magnetic contrast M(r) map obtained by subtracting STM topographs acquired over the same region of the sample at + 4 T and -4 T applied perpendicular to the sample surface (the minus sign denotes reversal of the magnetic field applied parallel or antiparallel to the sample c-axis). The thick white solid lines in (a) outline the rotational AF domain walls, across which the wave vector rotates by 90 degrees in-plane. Thinner white dashed lines denote smaller anti-phase sub-domains within each rotational domain, with the same wave vector but offset by π phase with respect to one another. (b) Antisymmetric strain map $U(r)$ calculated from an STM topograph over the region in (a) acquired at 0 T (Supplementary Note 1, Supplementary Figure 3, Supplementary Figure 6). (c,d) Amplitude of AF order along (c) the x-axis and (d) the y-axis determined from (a). Insets in (c,d) show azimuthally averaged cross-correlation coefficients between the image in (b) and images in (c,d), plotted as a function of distance. Box smoothing of about 1 nm length scale is applied in (b-d). The film was 24 nm thick and the measurement temperature was 4.5 K.

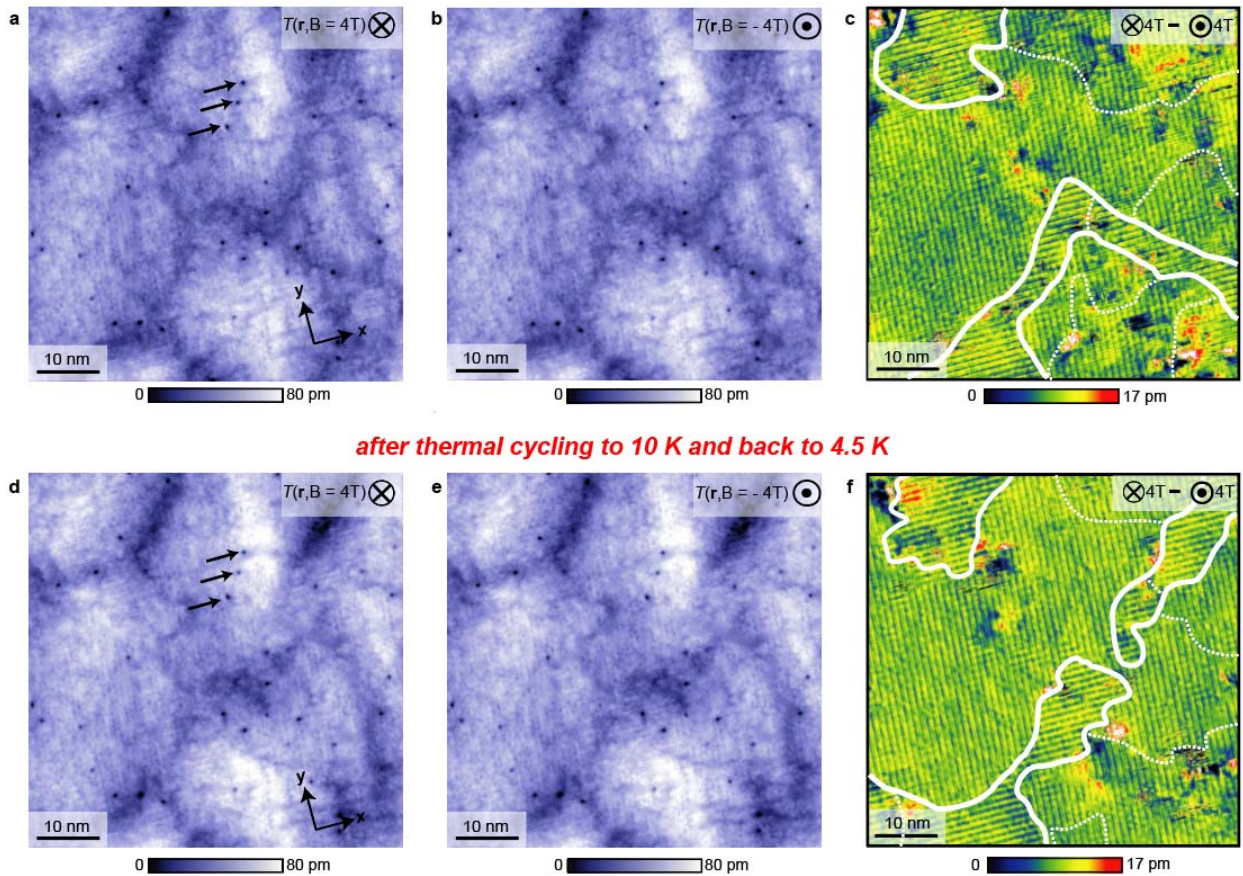

**Figure 4. Domain wall fluctuations with temperature change.** STM topograph of about 24 nm thick FeTe film acquired using a spin-polarized STM tip in (a) +4 T and (b) -4 T magnetic field (the minus sign denotes reversal of the magnetic field applied parallel or antiparallel to the sample c-axis). (c) Spin-resolved magnetic contrast $M(\mathbf{r})$ map, obtained by subtracting images in (a,b). The $M(\mathbf{r})$ map reveals the underlying AF order more clearly. The thick white solid lines in (c) outline the rotational AF domain walls, across which the wave vector rotates by about 90 degrees in-plane. Thinner white dashed lines denote smaller anti-phase sub-domains within each rotational domain, with the same wave vector but offset by π phase with respect to one another. (d-f) Equivalent panels to those in (a-c), over the same area of the sample, but after the sample was warmed up to 10 K and cooled back down. All STM topographs are acquired at 4.5 K on the same area of the sample surface. The arrows in (a,d) point to the same set of defects in both images to demonstrate the identical positions of the scan frames. The domain walls clearly shift after a thermal cycle, demonstrating their sensitivity to thermal fluctuations. STM setup conditions: (a-b, d-e) $I_{set}$ = 300 pA, $V_{sample}$ = 100 mV.